\documentclass[12pt,twoside]{article}
\usepackage{a4wide}
\usepackage{amssymb}
\usepackage{amsmath}
\usepackage{rotating}
\usepackage{color}
\usepackage{afterpage}
\usepackage{palatino}
\usepackage[pdftex]{hyperref}

\hypersetup{backref,colorlinks=false}

\makeatother

\begin{document}

\pdfinfo{/Title (The Tight-Binding method:application to AB s-valent dimer) /Author (D.G. Pettifor)}

\title{{\bf The Tight-Binding method: application to AB s-valent dimer}}
\author{D.G. Pettifor \\ Department of Materials \\ University of Oxford \\ Oxford OX1 3PH}
\date{(January 2010)}

\maketitle
\thispagestyle{empty}
\abstract{The AB s-valent dimer is used to analyse bond formation and charge transfer within the tight-binding (TB) approximation. In this way a physical
interpretation of the electronic structure and binding energy within density functional theory (DFT) is obtained which lends itself to the derivation of
covalent and ionic interatomic potentials. }
\clearpage
\tableofcontents

\pagenumbering{roman}
\clearpage

\pagenumbering{arabic}
%\section{}

\section {Introduction}

The AB $s$-valent dimer provides  the simplest system to illustrate the underlying approximations and concepts behind the TB method.  Whereas density functional theory (DFT) centres on the observable quantity $\rho \pmb {(r)}$, the electron density at all points in space $\pmb {r}$, the TB method coarse-grains the problem in terms of chemically-intuitive but non-unique quantities such as the bond integral $\beta \pmb {(R)}$, and overlap integral $S \pmb {(R)}$, defined in terms of the internuclear distance $\pmb {R}$ between a given pair of atoms.  In the following sections we will see how the TB method allows the DFT electronic structure and binding energy curves to be interpreted in terms of physical concepts such as covalency and ionicity. This leads naturally to a further coarse-graining of the problem in terms of covalent and ionic interatomic potentials.

\section {Electronic structure}

\subsection{Energy levels}

Let us consider bringing together two $s$-valent atoms A and B to form the AB diatomic molecule, as illustrated in Fig. 1 [1]. This could represent, for example, either the homovalent dimer $H_2$ (with $1s$ valence orbitals interacting) or the heterovalent dimer LiH (with the $2s$ Li valence orbital interacting with the $1s$ H orbital). The resultant electronic structure of the dimer is obtained by solving the corresponding Schr\"odinger equation

\begin{equation}
- \frac {1} {2} \nabla^2 \psi_{AB} \pmb {(r)} + V_{AB} \pmb {(r)} \psi_{AB} \pmb {(r)}= E \psi_{AB} \pmb {(r)} \,,
\end{equation}

\begin{figure} [b!]
\center
\includegraphics[width=.8\textwidth]{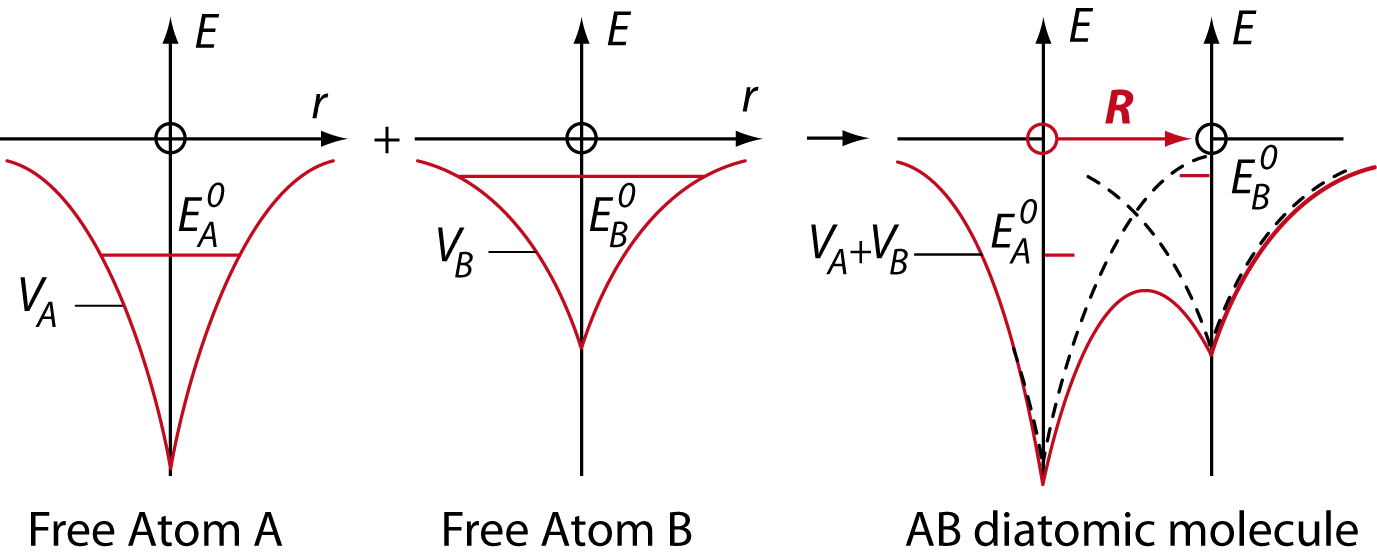}
\caption{The potential of the AB diatomic molecule may be approximated by the overlapping of the free atom A and B potentials. $E_A^0$ and $E_B^0$ are the free-atom energy levels of the valence s electrons. The internuclear distance $\pmb {R}$ runs from A to B.} \label{fig:AB_diatomic_molecule} 
\end{figure}

\mbox {} \\ where we have used atomic units
\begin{equation}
\hbar = 1, m = 1,e = 1,4 \pi \varepsilon_0 = 1\,.
\end{equation}
Thus, $\hbar^2 / 2m = \frac {1} {2}$ in Eq. (1) and the unit of energy is the Hartree (= 2 Rydberg = 27.2eV) and the unit of length is the au = 0.529\AA.

\mbox {} \\ The TB method solves this differential equation by assuming a minimal basis of atomic orbitals, namely

\begin{equation}
\psi_{AB} \pmb {(r)} = c_A \phi_A \pmb {(r)} + c_B \phi_B \pmb {(r-R)}
\end{equation}
where $\pmb {R}$ is the internuclear separation with the coordinate system centred on atom A as in Fig. 1.  The atomic orbitals satisfy the free-atom Schr\"odinger equations

\begin{align}
- \frac {1} {2} \nabla^2 \phi_A + V_A \phi_A &= E^0_A \phi_A \\
- \frac {1} {2} \nabla^2 \phi_B + V_B \phi_B &= E^0_B \phi_B 
\end{align}
with $E^0_A$ and $E^0_B$ being the free-atom energy levels of the valence $s$ electrons.  The coefficients $c_A$ and $c_B$ can be found by pre-multiplying the Schr\"odinger equation for the dimer Eq. (1) by $\psi_{AB} \pmb {(r)}$ , integrating over all space, and writing the resultant expression in the form

\begin{equation}
\int \psi_{AB} \left ( \hat{H} - E \right ) \psi_{AB} d \pmb {r} = 0 \,,
\end{equation}
where $\hat{H}$ is the Hamiltonian operator for the AB dimer, namely
\begin{equation}
\hat{H} = - \frac {1} {2} \nabla^2 + V_{AB} \,.
\end{equation}
Then substituting Eq. (3) into Eq. (6) we recover the TB {\it secular equation}

\begin{equation}
\left( 
\begin{array}{c}
H_{AA}-E \qquad H_{AB}-ES \\
H_{BA}-ES \qquad H_{BB}-E
\end{array} 
\right)
\left( 
\begin{array}{c}
c_A \\
c_B
\end{array} 
\right)
= 0 \,,
\end{equation}
where the {\it on-site} Hamiltonian matrix element $H_{AA}$ is given by

\begin{equation}
H_{AA} = \int \phi_A  \hat{H} \phi_A d \pmb {r} \,,
\end{equation}
and similarly for $H_{BB}$. The {\it intersite} Hamiltonian matrix is given by

\begin{equation}
H_{AB} = \int \phi_A \hat{H} \phi_B d \pmb {r} \,,
\end{equation}
and the {\it overlap integral} by

\begin{equation}
S = \int \phi_A \phi_B d \pmb {r} \,.
\end{equation}
The {\it orthogonal} TB model assumes that $S = 0$.  Within Molecular Orbital theory Eq. (8) is referred to as the H\"uckel equation for $S = 0$ and the extended H\"uckel equation for $S \neq 0$ [2].

\mbox {} \\ We will solve this TB secular equation for the eigenvalues and eigenvectors by working with respect to the average on-site energy of the dimer, namely

\begin{equation}
\bar {E} = \frac {1} {2} (H_{BB} + H_{AA}) \,.
\end{equation}
Then, defining

\begin{equation}
\varepsilon = E - \bar {E}
\end{equation}
and
\begin{equation}
\Delta = H_{BB} - H_{AA} \,,
\end{equation}
Eq. (8) takes the form

\begin{equation}
\left( 
\begin{array}{c}
- \frac {\Delta} {2} - \varepsilon \qquad \beta - \varepsilon S \\
\beta - \varepsilon S \qquad \frac {\Delta} {2} - \varepsilon
\end{array} 
\right)
\left( 
\begin{array}{c}
c_A \\
c_B
\end{array} 
\right)
= 0 \,.
\end{equation}
This has non-trivial solutions for

\begin{equation}
\left| 
\begin{array}{c}
- \frac {\Delta} {2} - \varepsilon \qquad \beta - \varepsilon S \\
\beta - \varepsilon S \qquad \frac {\Delta} {2} - \varepsilon
\end{array} 
\right|
= 0 \,,
\end{equation}
where the bond integral $\beta$ is given by

\begin{equation}
\beta = H_{AB} - \bar{E} S \,.
\end{equation}
We see that the {\it absolute} energy E with respect to the vacuum level does not enter Eq. (16) explicitly, only the {\it relative} energy $\varepsilon = (E - \bar {E})$.  The absolute energy $\bar {E}$ is hidden within the definition of $\beta$; the physical importance of this grouping will be discussed in section 2.3. 

\mbox {} \\ The {\it exact} eigenvalues of the TB determinantal Eq. (16) are given by

\begin{equation}
\varepsilon^{\pm} = \left [- \beta S \mp \sqrt {\beta^2 + (1-S^2) (\Delta/2)^2} \right ] / (1-S^2) \,.
\end{equation}
In order to keep our analytic analysis of this TB model for the dimer tractable for $\Delta \neq 0$, we now neglect all second and higher order terms in the overlap $S$ since they only lead to third-order and higher contributions such as $\beta S^2$ and $\Delta S^2$ in the eigenspectra. The eigenvalues then take the form 

\begin{equation}
E^{\pm} = \bar{E} + |\beta|S \mp \sqrt {\beta^2 + (\Delta/2)^2} \,,
\end{equation}
since the bond integral $\beta$ will be seen in section 2.3 to be negative.  Thus, as illustrated in the left-hand panel of Fig. 2, the eigenvalues comprise a bonding state and an anti-bonding state that are separated by $2|\beta|$ and $2|\beta_{AB}|$ for the homovalent and heterovalent cases respectively, where $\beta_{AB} = \sqrt {\beta^2 + (\Delta/2)^2}$. The shift in the energies due to the overlap repulsion $|\beta|S$ is not shown explicitly in Fig. 2.

\begin{figure} [h!]
\center
\includegraphics[width=.6\textwidth]{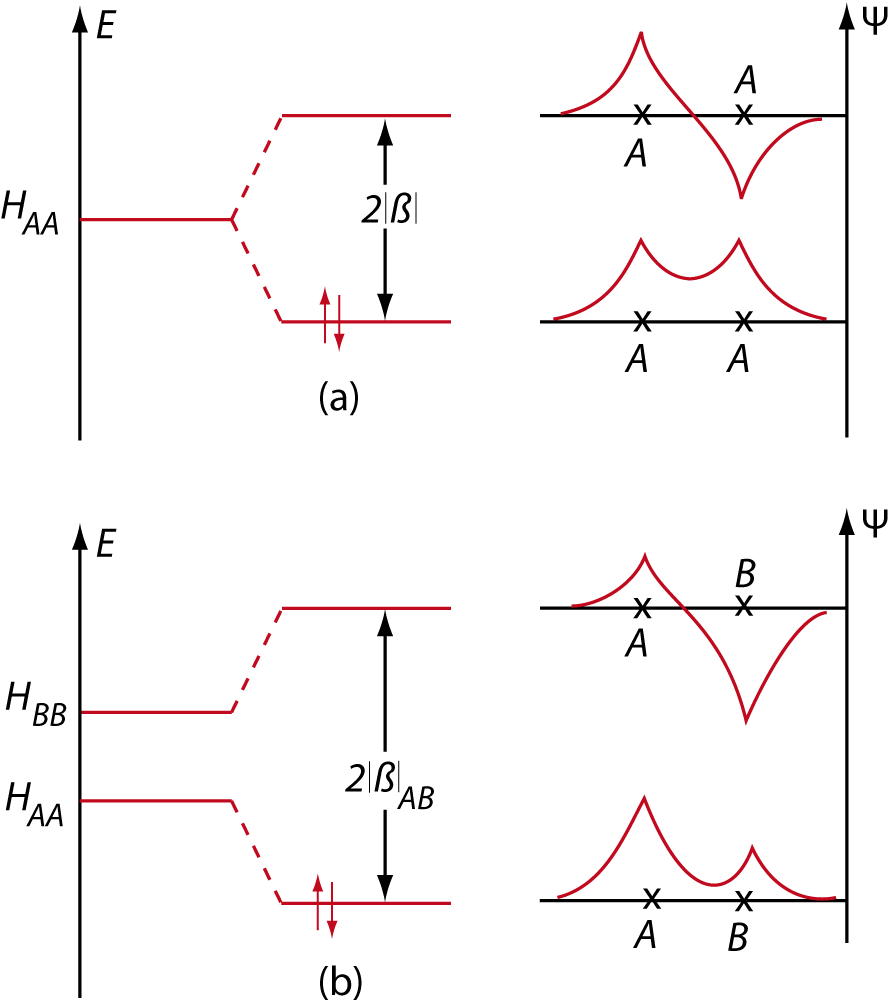}
\caption{The bonding and antibonding states for (a) the homonuclear and (b) the heteronuclear diatomic molecule. The shift in the energy levels due to overlap repulsion has not been shown.} \label{fig:bonding_antibonding} 
\end{figure}

\subsection{Charge density}
As illustrated schematically in Fig. 2, $H_2$ and $LiH$ have their two valence electrons occupying the bonding state $\psi^+_{AB} (\pmb {r})$ with anti-parallel spins.  Substituting the eigenvalues $E^{\pm}$ from Eq. (19) into the TB secular equation we find that the eigenfunctions are given by 

\begin{equation}
\psi^{\pm}_{AB} (\pmb {r}) = c^{\pm}_A \phi_A (\pmb {r}) + c^{\pm}_B \phi_B (\pmb {r-R}) \,,
\end{equation}
where

\begin{equation}
c^{\pm}_A = \frac {1} {\sqrt{2}} \left [1 \pm (\hat{\Delta} - S) / \sqrt {1 + \hat{\Delta}^2} \right ]^{\frac {1} {2}}
\end{equation}
and

\begin{equation}
c^{\pm}_B = \frac {1} {\sqrt{2}} \left [1 \mp (\hat{\Delta} + S) / \sqrt {1 + \hat{\Delta}^2} \right ]^{\frac {1} {2}} \,,
\end{equation}
neglecting second and higher order contributions in the overlap.  The normalized atomic energy-level mismatch $\hat{\Delta}$ is defined by

\begin{equation}
\hat{\Delta} = \Delta / (2|\beta|) \,.
\end{equation}
Thus, for a given internuclear separation $\pmb {R}$ the {\it number density} of the valence electrons in the AB dimer can be written as

\begin{equation}
\rho_{AB} (\pmb {r}) = 2[ \psi^+_{AB} (\pmb {r})]^2 \,,
\end{equation}
with the corresponding {\it electronic-charge density} $=-e \rho_{AB} (\pmb {r})= -\rho_{AB} (\pmb {r})$  in atomic units.  It follows from Eqs. (20) and (24) that 

\begin{align}
\rho_{AB} (\pmb {r}) &= 2 (c^+_A)^2 \rho_A (\pmb {r}) + 2 (c^+_B)^2 \rho_B (\pmb {r} - \pmb {R}) + 4 c^+_A c^+_B \phi_A (\pmb {r}) \phi_B (\pmb {r} - \pmb {R}) \\ \nonumber
&= N_A \rho_A (\pmb {r}) + N_B \rho_B (\pmb {r} - \pmb {R}) + 2 \Theta_{AB} \phi_A (\pmb {r}) \phi_B (\pmb {r} - \pmb {R}) \,,
\end{align}
where

\begin{equation}
\rho_{A(B)} (\pmb {r}) = [ \phi_{A(B)} (\pmb {r})]^2 \,.
\end{equation}
$N_A$ and $N_B$ are the number of valence electrons on atoms A and B respectively, namely

\begin{equation}
N_A = 1 + \hat{\Delta} / \sqrt {1 + \hat{\Delta}^2} - S / \sqrt {1 + \hat{\Delta}^2} \,,
\end{equation}
and

\begin{equation}
N_B = 1 - \hat{\Delta} / \sqrt {1 + \hat{\Delta}^2} - S / \sqrt {1 + \hat{\Delta}^2} \,,
\end{equation}
to first order in $S$. The prefactor $\Theta_{AB}$ in the last term of Eq. (25) is the bond order between atoms A and B, namely

\begin{equation}
\Theta_{AB} = 1 / \sqrt {1 + \hat{\Delta}^2} - S \,.
\end{equation}
Eq. (25) for the number density may be simplified by grouping the overlap contribution in the first two terms with the bond-order contribution in the last term to give 

\begin{equation}
\rho_{AB} (\pmb {r}) = (1+q) \rho_A (\pmb {r}) + (1-q) \rho_B (\pmb {r} - \pmb {R}) + \Theta \rho_{cov} (\pmb {r}) \,,
\end{equation}
where

\begin{equation}
\rho_{cov} (\pmb {r}) = 2 \phi_A (\pmb {r}) \phi_B (\pmb {r} - \pmb {R}) - S [\rho_A (\pmb {r}) + \rho_B (\pmb {r})] \,.
\end{equation}
It follows from Eqs. (27) and (28) that the {\it charge} $q$ is given by

\begin{equation}
q = \hat{\Delta} / \sqrt {1 + \hat{\Delta}^2} \geq 0 \,,
\end{equation}
which is positive due to our choice of $E_B \geq E_A$  in Fig. 1.  On the other hand, it follows from Eq. (29) that the {\it bond order} $\Theta$  is given by 

\begin{equation}
\Theta = 1 / \sqrt {1 + \hat{\Delta}^2} = \sqrt {1 - q^2} \,.
\end{equation}
We see from Eq. (32) that the normalized atomic energy-level mismatch $\hat{\Delta}$  can be written in terms of the charge $q$ as

\begin{equation}
\hat{\Delta} = q /\sqrt {1 - q^2} \,.
\end{equation}
Thus, $\hat{\Delta} \rightarrow 0$  corresponds to the {\it covalent} limit with $q=0$  and $\Theta=1$ , whereas $\hat{\Delta} \rightarrow \infty$  corresponds to the {\it ionic} limit with $q=1$ and $\Theta=0$. 

\mbox {} \\ The nature of the covalent-bond density $\rho_{cov} (\pmb {r})$ is illustrated in Fig. 3 for the hydrogen molecule where we see, as expected, that the electrons flow from outside the bond region to inside.  Importantly, the total {\it covalent-bond charge} is zero because

\begin{equation}
\int \rho_{cov} (\pmb {r}) d \pmb {r} = 2S - 2S = 0\,.
\end{equation}
Interestingly, the net {\it atomic charge} of $\pm q$ is consistent with the definition of Mulliken charge where the overlap contribution $2 \Theta_{AB}S$ from Eq. (25) is assumed to be divided evenly between both atoms A and B [2].  It follows from Eqs. (27), (28) and (29) with the values of $q$ and $\Theta$ from Eqs. (32) and (33) that to first order in $S$

\begin{equation}
q_{Mulliken} = (q - \Theta S) + \Theta S = q\,.
\end{equation}
In this section we have demonstrated that a given choice of $\Delta$  and $\beta$  in the TB secular equation leads to a specific value for the {\it atomic charge} $q$ through Eq. (32). In the next section we will show that these TB parameters $\Delta$ and $\beta$ are themselves functions of the charge so that Eq. (32) must be solved self-consistently for the charge $q$.

\begin{figure} 
\center
\includegraphics[width=.4\textwidth]{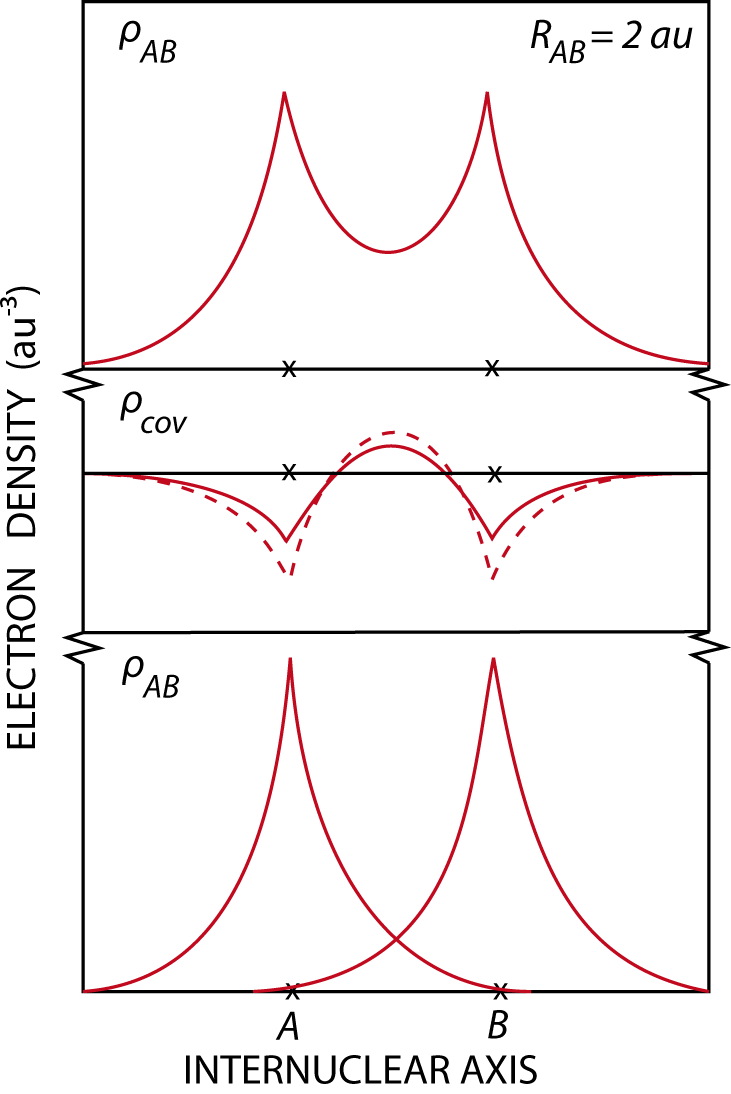}
\caption{The electron density of the homonuclear molecule (upper panel) can be regarded as the sum of the non-interacting or frozen free-atom electron densities (lower panel) and the quantum mechanically induced covalent-bond density (middle panel).  The dashed curve represents the first-order approximation, Eq. (31), for the bond density, the deviation from the exact result (full curve) being due to the sizeable value of the overlap integral namely $S = 0.59$ at $R = 2 au$.} \label{fig:electron_density} 
\end{figure}

\subsection{Expressions for TB parameters}  

The key parameters that enter the TB expression for the energy levels, namely the atomic energy-level mismatch $\Delta$ and the bond-integral $\beta$, depend on the Hamiltonian matrix elements $H_{AA}$, $H_{BB}$ and $H_{AB}$ through Eqs. (14) and (17). These matrix elements in their turn depend on the potential $V_{AB} (\pmb {r})$ through Eqs. (7), (9) and (10).  Within DFT the potential seen by the electron in the Kohn-Sham equations can be written as the sum of three contributions

\begin{equation}
V (\pmb {r}) = V_{ion} (\pmb {r}) + V_H (\pmb {r}) + V_{xc} (\pmb {r})\,,
\end{equation}
where the first is the potential due to the ion cores, the second the Coulomb or Hartree potential from the valence charge density, and the third the exchange-correlation potential that enters within DFT.  In this analytic treatment of the dimer we will neglect this latter contribution as it varies non-linearly with the density unlike the Coulomb term.  We will assume that it can be subsumed into the TB parameters when they are fitted to the DFT eigenspectra and binding energy curves.  We will see in section 3.3 that this is indeed a good approximation as it leads to a consistent physical picture for both the non-magnetic DFT energy levels and the binding-energy curves of the hydrogen molecule.

\mbox {} \\ Thus, using Eq. (30) for the valence density and neglecting the exchange-correlation contribution, the potential $V_{AB} (\pmb {r})$ for the AB dimer takes the form 

\begin{equation}
V_{AB} (\pmb {r}) = V_A (\pmb {r}) + V_B (\pmb {r}) - q [V_B^{\rho} (\pmb {r}) - V_A^{\rho} (\pmb {r})] + \sqrt {1 - q^2} V_{cov}^{\rho} (\pmb {r}) \,.
\end{equation}
The first two terms are the Coulomb potential that results from overlapping neutral free-atomic potentials, as sketched on the right-hand side of Fig. 1.  That is, 

\begin{equation}
V_{A(B)} (\pmb {r}) = V_{A(B)}^{ion} (\pmb {r}) + V_{A(B)}^{\rho} (\pmb {r}) \,,
\end{equation}
where $V_{ion}$ is the potential for the ion core that falls off inversely with distance outside the core region such that

\begin{equation}
V_{A(B)}^{ion} (\pmb {r}) = Z_{A(B)} / \left |\pmb {r} - \pmb {R}_{A(B)} \right | \text {for} \left |\pmb {r} - \pmb {R}_{A(B)} \right | \geq \pmb {R}_{A(B)}^{core} \,.
\end{equation}
For our monovalent AB dimer $Z_A = Z_B = 1$ and $\pmb {R}_A = 0, \pmb {R}_B = \pmb {R}$ from Fig. 1.  Whereas for hydrogen the core radius is zero, for lithium it takes some finite value enclosing the $1s$ shell of core electrons. $V^{\rho}$ is the potential due to the valence electrons of the free atoms, namely

\begin{equation}
V_{A(B)}^{\rho} (\pmb {r}) = \int \rho_{A(B)} (\pmb {r'} - \pmb {R}_{A(B)}) / | \pmb {r} - \pmb {r'}| d \pmb {r'} \,.
\end{equation}
The superscripts $\rho$ in Eqs. (39) and (41) are used to remind us that these potential contributions arise solely from the valence electrons.

\mbox {} \\ The {\it input} potential for most TB calculations is assumed to be the sum of neutral atomic potentials $(V_A + V_B)$ as in Fig. 1.  However, as can be seen from Eq. (38) the {\it output} potential can differ from this input potential due to the flow of charge.  In particular, the third contribution in Eq. (38) results from the shift in the potential due to the flow of charge $q$ from one atom to another, whereas the fourth contribution arises from the creation of the covalent bond with a bond order $\sqrt {1-q^2}$. This latter term is given by

\begin{equation}
V_{cov}^{\rho} (\pmb {r}) = \int \rho_{cov} (\pmb {r'}) / | \pmb {r} - \pmb {r'}| d \pmb {r'} \,.
\end{equation}
Thus, for homovalent dimers such as $H_2$ with $\Delta = 0$ the output potential simplifies to 
\begin{equation}
[V_{AB} (\pmb {r})]_{\Delta=0} = V_A (\pmb {r}) + V_B (\pmb {r}) + V_{cov}^{\rho} (\pmb {r})\,.
\end{equation}
A {\it self-consistent} TB calculation requires the output potential to be identical to the input potential.  This will be discussed in the next section once we have determined explicit expressions for the on-site and intersite Hamiltonian matrix elements, and hence the appropriate TB parameters. 

\mbox {} \\ The {\it on-site} Hamiltonian matrix elements $H_{AA}$ and $H_{BB}$ follow from Eqs. (7), (9) and (38), namely 

\begin{equation}
H_{AA} = E_A^0 + \alpha_A + (J_{AA} - J_{AB}) q + J_{Ac} \sqrt {1 - q^2}\,,
\end{equation}
where $E_A^0$ is the atomic energy level of the isolated free atom.  The second contribution is the shift in the on-site energy on atom A due to the crystal field from neighbouring atom B, that is

\begin{equation}
\alpha_A = \int \rho_A V_B d \pmb {r}\,.
\end{equation}
The third contribution is the additional shift in the on-site energy on atom A due to the flow of charge $q$, which gives rise to an upward shift $J_{AA} q$ due to the increased number $+q$ of electrons on site A and a downward shift $J_{AB} (-q)$ due to the Coulomb attraction resulting from the negative charge -q on the neighbouring site B.  

\mbox {} \\ The {\it on-site Coulomb integral} (traditionally denoted by $U$ in the many-body Hubbard Hamiltonian) is given by

\begin{equation}
J_{AA} = \int \int \rho_A (\pmb {r}) \rho_A (\pmb {r'}) / | \pmb {r} - \pmb {r'}| d \pmb {r} d \pmb {r'} \,
\end{equation}
and the {\it intersite Coulomb integral} by

\begin{equation}
J_{AB} (\pmb {R}) = \int \int \rho_A (\pmb {r}) \rho_B (\pmb {r' - R}) / | \pmb {r} - \pmb {r'}| d \pmb {r} d \pmb {r'} \,.
\end{equation}
The fourth contribution is the Coulomb shift in the on-site energy on atom A due to the formation of the covalent bond, where the {\it atom-covalent bond Coulomb integral} is given by 

\begin{equation}
J_{Ac} = \int \int \rho_A (\pmb {r}) \rho_{cov} (\pmb {r'}) / | \pmb {r} - \pmb {r'}| d \pmb {r} d \pmb {r'} \,.
\end{equation}
Similarly, the on-site energy on atom B is given by

\begin{equation}
H_{BB} = E_B^0 + \alpha_B - (J_{BB} - J_{AB}) q + J_{Bc} \sqrt {1 - q^2} \,.
\end{equation}
The {\it inter-site Hamiltonian matrix element} $H_{AB}$ follows from Eqs. (7), (10) and (38), namely
\begin{align}
H_{AB} = \beta_{SK} &+ \frac {1} {2} (E_B^0 + E_A^0)S - q \int \phi_A (V_B^{\rho} - V_A^{\rho}) \phi_B d \pmb {r} \nonumber \\ 
                    &- q \int \phi_A (V_B^{\rho} - V_A^{\rho}) \phi_B d \pmb {r} + \sqrt {1 - q^2} \int \phi_A V_{cov}^{\rho} \phi_B d \pmb {r}  \,
\end{align}
where the first term is the well-known Slater-Koster {\it two-centre bond (or hopping) integral} for orthogonal orbitals, namely
\begin{equation}
\beta_{SK} (\pmb {R}) = \int \phi_A (\pmb {r}) [(V_A + V_B)/2] \phi_B (\pmb {r-R}) d \pmb {r} \,.
\end{equation}
Because the valence $s$-orbitals $\phi_A$ and $\phi_B$ are angularly independent, this bond integral is a function only of $R$, the magnitude of the internuclear distance $R$, and not the direction $\hat {\pmb {R}}$. $\beta_{SK}$ is clearly {\it negative} for atomic $s$ orbitals interacting via the negative overlap of the atomic potentials (see Fig. 1). 

\mbox {} \\ Finally, using Eq. (31) to express the quantum-mechanical interference factor $\phi_A (\pmb {r}) \times$ $ \phi_B (\pmb {r-R})$ in terms of the overlap integral $S$ and the covalent-bond density $\rho_{cov}$, the intersite Hamiltonian matrix element can be written as
\begin{equation}
H_{AB} = \beta_{SK} + \frac {1} {2} (E_B^0 + E_A^0)S - \frac {1} {2} [(J_{Bc} - J_{Ac}) + (J_{BB} - J_{AA})S]q + \frac {1} {2} [J_{cc} + (J_{Bc} + J_{Ac}) S] \sqrt {1 - q^2}  \,,
\end{equation}
where the {\it covalent-bond-covalent-bond Coulomb integral} is given by

\begin{equation}
J_{cc} = \int \int \rho_{cov} (\pmb {r}) \rho_{cov} (\pmb {r'}) / | \pmb {r} - \pmb {r'}| d \pmb {r} d \pmb {r'}  \,.
\end{equation}
The key TB parameters $\bar {E}, \Delta$ and $\beta$  can now be found by substituting Eqs. (44) and (52) into Eqs. (12), (14) and (17) respectively.  

\mbox {} \\ We find the {\it average on-site energy} of the dimer is given by 

\begin{equation}
\bar {E} = \frac {1} {2} (H_{BB} + H_{AA}) = \bar {E}_0 + \frac {1} {2} (\alpha_B + \alpha_A) - \frac {1} {2} (J_{BB} - J_{AA}) q + \frac {1} {2} (J_{Bc} + J_{Ac}) \sqrt {1 - q^2} \,,
\end{equation}
where $\bar {E}_0 =\frac {1} {2} (E_B^0 + E_A^0)$.  

\mbox {} \\ On the other hand, the {\it atomic energy-level mismatch} in the dimer can be written

\begin{equation}
\Delta = H_{BB} - H_{AA} = \Delta_0 + (\alpha_B - \alpha_A) - 2Jq + (J_{Bc} - J_{Ac})\sqrt {1 - q^2} \,,
\end{equation}
where $\Delta_0 = (E_B^0 - E_A^0)$  and the effective Coulomb integral

\begin{equation}
J = \frac {1} {2} (J_{BB} + J_{AA}) - J_{AB} \,.
\end{equation}
Finally, the {\it bond integral} can be expressed in the form

\begin{equation}
\beta = H_{AB} - \bar {E} S = [\beta_0 - \frac {1} {2} (J_{Bc} - J_{Ac})q + \frac {1} {2} J_{cc} \sqrt {1 - q^2}]  \,,
\end{equation}
where

\begin{equation}
\beta_0 = \beta_{SK} - \frac {1} {2} (\alpha_A + \alpha_B)S \,.
\end{equation}
The first term in Eq. (57) is, therefore, the usual Slater-Koster bond integral that has been modified by the non-orthogonality of the atomic orbitals.  The second term is the contribution to the bond integral due to the creation of the point charges $\pm q$ , whereas the third term arises from the formation of the bond-charge density with bond order $\sqrt {1 - q^2}$.

\subsection{Self-consistency}

We have seen from the discussion following Eq. (38) that the {\it output} potential differs from the {\it input} potential due to the redistribution or flow of charge reflecting the formation of both the covalent bond with bond order $\sqrt {1 - q^2}$ and the ionic bond with point charges $\pm q$.  This in turn influences the values of the TB parameters.  In particular, the atomic energy-level mismatch $\Delta$ depends explicitly on the charge $q$ through Eq. (55),

\begin{equation}
\Delta (q) = \Delta_0 + (\alpha_B - \alpha_A) - 2Jq + (J_{Bc} - J_{Ac}) \sqrt {1 - q^2} \,.
\end{equation}
As expected, the right-hand side of this equation involves various Coulomb integrals that drive the shift in the on-site energy levels. On the other hand, we have already found by solving the TB secular equation that $\Delta$ depends explicitly on $q$ through the simple relationship of Eq. (34), namely

\begin{equation}
\Delta (q) = 2|\beta|\hat{\Delta} = 2|\beta|q / \sqrt {1 - q^2} \,.
\end{equation}
Thus, substituting Eq. (57) for $\beta$ into Eq. (60) and equating the right-hand sides of Eqs. (59) and (60), we find the following self-consistency equation for q, namely

\begin{equation}
q = \frac {\Delta_0 + (\alpha_B - \alpha_A) + (J_{Bc} - J_{Ac}) (1 - 2 q^2) / \sqrt {1 - q^2}} {2 J - J_{cc} + 2 |\beta_0| / \sqrt {1 - q^2}}\,.
\end{equation}
Within traditional TB methods the contributions arising from the {\it differences} in the crystal-fields and atom-bond Coulomb integrals are neglected in the numerator, and the {\it small} bond-bond Coulomb integral is neglected in the denominator. The simplified self-consistency equation then takes the form
\begin{equation}
q = \frac {\Delta_0} {2 J + 2 |\beta_0| / \sqrt {1 - q^2}} =  \frac {\tilde{\Delta}_0} {1 + 2 |\tilde{\beta}_0| / \sqrt {1 - q^2}}\,,
\end{equation}

\mbox {} \\ where $\tilde{\Delta}_0$ and $\tilde{\beta}_0$ have been normalized by $2J$ i.e.
\begin{equation}
\tilde{\Delta}_0 = \Delta_0 /2J , \tilde{\beta}_0 = \beta_0 /2J  \,.
\end{equation}

\mbox {} \\ This corresponds to the quartic equation
\begin{equation}
q^4 = - 2 \tilde{\Delta}_0 q^3 + [4 \tilde{\beta}_0^2 +  \tilde{\Delta}_0^2 - 1] q^2 + 2 \tilde{\Delta}_0 q - \tilde{\Delta}_0^2 = 0 \,.
\end{equation}

\begin{figure} 
\center
\includegraphics[width=.6\textwidth]{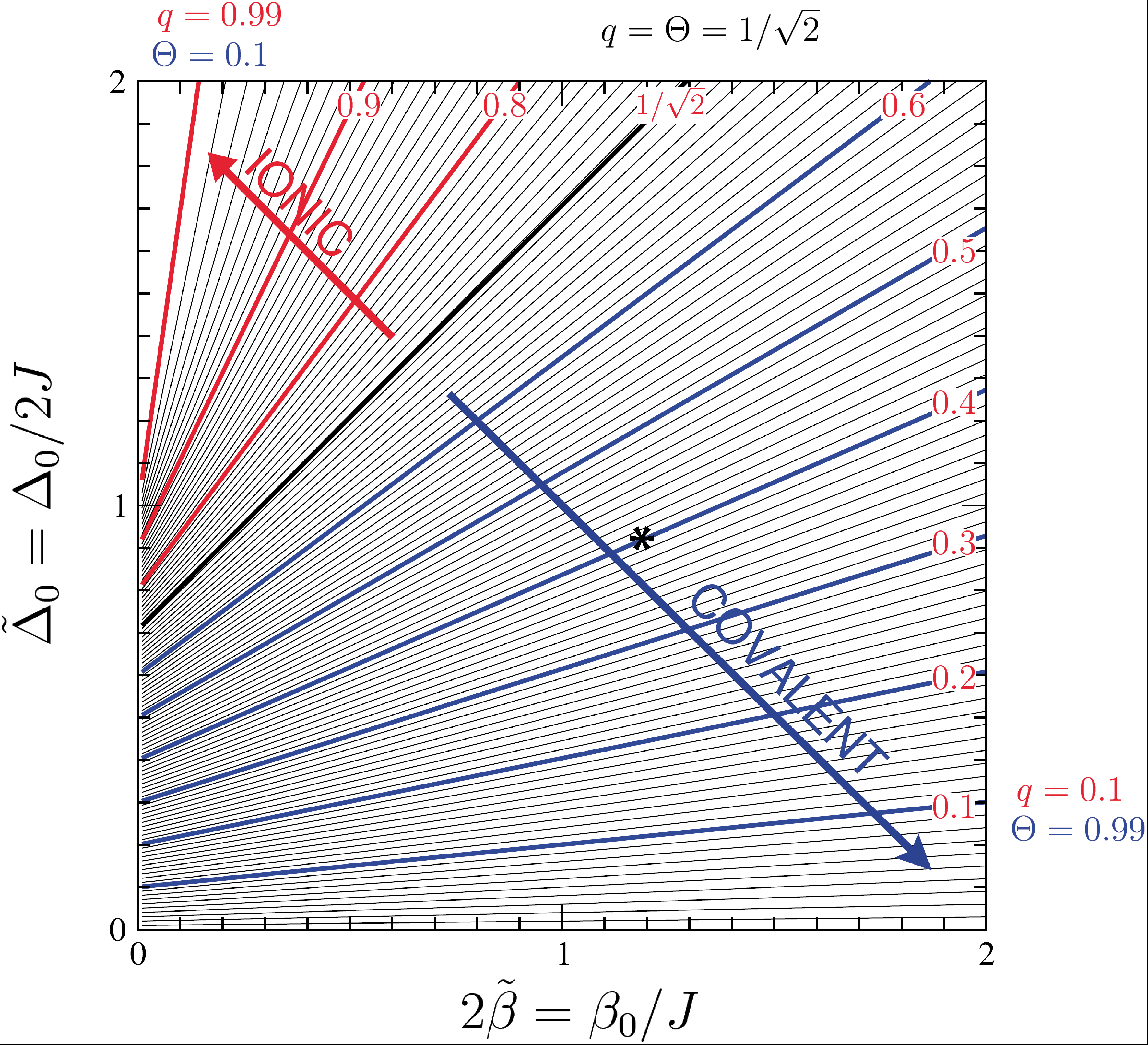}
\caption{Contours of self-consistent charge $q$ (and corresponding bond order $\Theta = \sqrt {1 - q^2})$ as a function of $2 \tilde{\beta}_O$ and $\tilde{\Delta}_O$. Solid black curve corresponds to $q = \Theta = 1 / \sqrt {2}$  when degree of ionicity equals degree for covalency.  Coordinate point for LiH, marked by *, corresponds to (1.2, 0.88) in $q=0.4$. \label{fig:charge_q}} 
\end{figure}

\mbox {} \\ Fig. 4 shows the contours of the self-consistent charge $q$ (and corresponding bond order $\Theta$) as a function of the two variables $2 \tilde{\beta}_0$ and $\tilde{\Delta}_0$ , where we have chosen the co-ordinates $(2 \tilde{\beta}_0, \tilde{\Delta}_0)$  in keeping with Eqs. (23)and (62). It follows from Eq. (63) that
\begin{equation}
\tilde{\Delta}_0 = (q / \sqrt {1 - q^2}) 2 \tilde{\beta}_0 + q  \,,
\end{equation}
so that the curves in Fig. 4 for different q vary linearly, as observed.
The solid curve corresponds to $q = \Theta = 1 / \sqrt {2}$  when the charge and bond-order take equal values.  Since the maximum charge and bond-order for an $s$-valent dimer is $1$ we can regard $q$ as the degree of ionicity of the bond and $\Theta$  as the degree of covalency.  Thus, the ionicity and covalency increase in the direction of the arrows shown.  As expected, for homovalent dimers with the normalized atomic energy-level mismatch $\tilde{\Delta}_0 = 0$  we have fully covalent dimers with $q = 0, \Theta = 1$. For the normalized bond integral $\tilde{\beta}_0 = 0$ corresponding to isolated free atoms we have from Eq. (62) that

\begin{equation}
q =
\begin{cases}
\tilde{\Delta}_0 \quad &for \quad \tilde{\Delta}_0 \leq 1 \\
1 \quad &for \quad \tilde{\Delta}_0 > 1
\end{cases} 
\,.
\end{equation}
Therefore, the different constant $q$-curves in Fig. 4 intercept the vertical axis at $\tilde{\Delta}_0 = q$ for $q < 1$. We will return to a discussion of the free atom case and the heterovalent dimer LiH in the next section once we have found explicit values for the TB parameters $\tilde{\Delta}_0$ , $\tilde{\beta}_0$ and $J$.

\section {Binding energy}
\subsection {Total energy}

The {\it total} energy of the AB $s$-valent dimer can be written in the form

\begin{equation}
U = U_{band} - U_{dc} + 1/R  \,,
\end{equation}
where the first contribution is the band energy and the second term is the double-counting energy.  The third contribution is the core-core repulsive energy (for the case of LiH this only holds for internuclear separations $R>R_{Li}^{core}$ (c.f. Eq. (40)). We have neglected the usual exchange-correlation contributions, assuming that they have been absorbed implicitly within the {\it fitted} on-site and inter-site TB parameters. 

\mbox {} \\ The {\it band} energy is the sum over the occupied eigenvalues, namely

\begin{equation}
U_{band} = \sum_{n \text{ }occ} 2 E_n  
\end{equation}
where the prefactor $2$ accounts for the spin-degeneracy in non-magnetic systems (c.f. Eqs. (1.175) and (7.13) of [3]).  Note that we have retained the nomenclature 'band' for this sum over occupied eigenvalues even though molecules have a {\it discrete} spectrum rather than a band of states as in bulk materials.  

\mbox {} \\ For the AB dimer 

\begin{equation}
U_{band} = 2 E^+ = 2 [|\beta|S + \frac {1} {2} (H_{AA} + H_{BB}) - |\beta| / \sqrt {1 - q^2}] \,, 
\end{equation}
where from Eq (33) we have substituted

\begin{equation}
\sqrt {1 + \hat {\Delta}^2} = 1 / \sqrt {1 - q^2} \,, 
\end{equation}
into the eigenvalue in Eq. (19).  The first term represents the upward shift in the eigenvalue due to the overlap repulsion.  The second term locates the {\it absolute energy} of the average on-site energies with respect to the vacuum level, and the third term represents the downward shift in the eigenvalue due to the formation of the covalent bond. 

\mbox {} \\ A more transparent understanding of these last two contributions can be obtained by rewriting them as

\begin{equation}
(H_{AA} + H_{BB}) - 2 |\beta| / \sqrt {1 - q^2} =
\left\{ 
\begin{array}{c}
(1 + q) H_{AA} + (1 - q) H_{BB}  \\
+ q (H_{BB} - H_{AA}) - 2 |\beta| / \sqrt {1 - q^2}
\end{array} 
\right\}
\,.
\end{equation}
But from Eqs. (14) and (60)

\begin{equation}
(H_{BB} - H_{AA}) = \Delta = 2 |\beta|\hat {\Delta} = 2 |\beta| q / \sqrt {1 - q^2}] \,. 
\end{equation}
Therefore, the band energy can be re-expressed as

\begin{equation}
U_{band} = 2 |\beta| S + (1 + q) H_{AA} + (1 - q) H_{BB} - 2 |\beta| \sqrt {1 - q^2}] \,,
\end{equation}
where from Eq. (33) $\sqrt {1 - q^2}$ is the bond order $\Theta$ . Thus, the {\it band} energy is given by the sum of the overlap repulsion, the energy of $(1 + q)$ and $(1 - q)$ electrons sitting on the A and B sites respectively, and the {\it bond} energy (given by the product of the bond order and the bond integral).

\mbox {} \\ This physically transparent result for the dimer is a specific example of the well-known general result that 

\begin{equation}
U_{band} = \sum_{nocc} 2 E_n = \text {Tr } \hat {\rho} \hat {H} \,,
\end{equation}
where $\hat {\rho}$ and $\hat {H}$  are the density and Hamiltonian operators respectively (see, for example, [3]).  Expanding the eigenfunctions in terms of local orbitals as in Eq. (3) leads to 

\begin{equation}
\text {Tr } \hat {\rho} \hat {H} = 2 \sum_{nocc} \sum_{i,j} c_i^n c_j^n H_{ji} \,,
\end{equation}
assuming real eigenvectors.  Thus, for our AB dimer

\begin{equation}
U_{band} = 2 c_A^2 H_{AA} + 2 c_B^2 H_{BB} + 4 c_A c_B H_{AB} \,.
\end{equation}
Taking the eigenvectors from Eqs. (21) and (22) we can write

\begin{equation}
U_{band} =
\left\{ 
\begin{array}{l}
[(1 + q) - S \sqrt {1 - q^2}] H_{AA} + [(1 - q) - S \sqrt {1 - q^2}] H_{BB}  \\
+ 2 [\sqrt {1 - q^2} - S] H_{AB} 
\end{array} 
\right\}
\end{equation}
where we have used Eqs. (32) and (33).  Finally, substituting $H_{AB} = \beta + \frac {1} {2} (H_{AA} + H_{BB}) S$  from Eq. (17) into the above equation, we recover

\begin{equation}
U_{band} = [(1 + q) H_{AA} + (1 - q) H_{BB} - 2 |\beta| \sqrt {1 - q^2} + 2 |\beta| S] 
\end{equation}
to first order in $S$, where the repulsive overlap contribution arises from the non-orthogonality term in the original bond order $\Theta_{AB}$ of Eq. (29).  Thus, Eq. (78) obtained from the trace of $\hat {\rho} \hat {H}$  is identical to Eq. (73) obtained from the eigenvalues directly, as it must.

\mbox {} \\ The {\it band} energy, Eq. (73), can be expressed explicitly in terms of the TB parameters by substituting in both Eq. (54) for the average on-site energy $\bar {E}$  and also Eq. (55) for the atomic energy-level mismatch $\Delta$.  It follows that

\begin{align}
U_{band} = 2 |\beta| S + (H_{BB} + H_{AA}) - q (H_{BB} - H_{AA}) - 2 |\beta| \sqrt {1 - q^2} \\ \nonumber
= \left\{ 
\begin{array}{l}
2 |\beta| S + (E_B^0 + E_A^0) + (\alpha_B + \alpha_A) - (J_{BB} - J_{AA})q \\
+ (J_{Bc} + J_{Ac}) \sqrt {1 - q^2} - q [\Delta_0 + (\alpha_B - \alpha_A)] \\
- 2 [|\beta_0| + \frac {1} {2} (\alpha_A + \alpha_B) S] \sqrt {1 - q^2} \\
+ J_{cc} (1 - q^2) + 2 J q^2
\end{array} 
\right\}
\end{align}
This band energy has double-counted the Coulomb energy between the electrons so that the total energy in Eq. (67) includes the double-counting correction term, namely

\begin{equation}
U_{dc} = \frac {1} {2} \int \int \rho (\pmb {r}) \rho (\pmb {r'}) / |\pmb {r} - \pmb {r'}| d \pmb {r} d \pmb {r'} \,.
\end{equation}
Substituting in the valence charge density from Eq. (30), this can be written in terms of the various Coulomb integrals as 

\begin{equation}
U_{dc} = 
\left\{ 
\begin{array}{l}
\frac {1} {2} J_{AA} (1 + q)^2 + \frac {1} {2} (1 - q)^2 J_{BB} + (1 - q^2) J_{AB}\\
+ \sqrt {1 - q^2} [(1 + q) J_{Ac} + (1 - q) J_{Bc}] + \frac {1} {2} (1 - q^2) J_{cc}
\end{array} 
\right\}
\,.
\end{equation}
The terms in the above double-counting expression can be regrouped to give

\begin{equation}
U_{dc} = 
\left\{ 
\begin{array}{l}
[\frac {1} {2} (J_{BB} + J_{AA}) + J_{AB}] + \frac {1} {2} J_{cc} \\
- \left [(J_{BB} - J_{AA}) q - (J_{Bc} + J_{Ac}) \sqrt {1 - q^2} \right ] \\
+ (J_{Bc} - J_{Ac}) q \sqrt {1 - q^2} + (J - \frac {1} {2} J_{cc}) q^2
\end{array} 
\right\}
\,.
\end{equation}
The first contribution is the electronic Coulomb energy of {\it neutral atoms} A and B assuming the atomic charge density remains frozen as the atoms are brought together from infinity to form the AB dimer.  The second contribution is the Coulomb self-energy of a {\it neutral covalent bond} that is fully saturated with a bond order of unity.  The next two terms in the square brackets reflect the {\it first order} shifts in the average on-site energy in Eq. (54) due to the charge q and bond order $\Theta = \sqrt {1 - q^2}$  correcting the dimer potential in Eq. (38) beyond that of overlapping neutral frozen atoms.  The following contribution is a cross-term that arises from the output charge q interacting with the covalent-bond potential in Eq. (38), and vice versa.  The last contribution in Eq. (82) is the {\it second-order} $q^2$ double-counting Coulomb terms. 

\mbox {} \\ The {\it total} energy of the AB dimer, Eq. (67), then takes the physically transparent form 

\begin{equation}
U (R,q) = [(U_B^0 + U_A^0) + U_{over} (R) + U_{es} (R) + U_{cov} (R,q) + U_{ionic} (R,q)]\,,
\end{equation}
where $(U_B^0 + U_A^0)$ is the energy of isolated neutral free atoms A and B, namely

\begin{equation}
\left.
\begin{array}{l}
U_A^0 = E_A^0 - \frac {1} {2} J _{AA} \\
U_B^0 = E_B^0 - \frac {1} {2} J _{BB} 
\end{array} 
\right\}
\,.
\end{equation}
The next four contributions on the right-hand side of Eq. (83) are the overlap, electrostatic, covalent and ionic terms respectively.  They are given to first order in the overlap S by

\begin{equation}
U_{over} (R) = 2|\beta_0|S \,,
\end{equation}
\begin{equation}
U_{es} (R) = (1/R + \alpha_B + \alpha_A - J_{AB}) + \frac {1} {2} J_{cc} \,,
\end{equation}
\begin{equation}
U_{cov} (R,q) = - 2|\beta_0| \sqrt {1 - q^2} \,,
\end{equation}
and

\begin{equation}
U_{ionic} (R, q) = 
\left\{ 
\begin{array}{c}
- [\Delta_O + (\alpha_B - \alpha_A)] q + (J_{Bc} - J_{Ac}) q \sqrt {1 - q^2}  \\
+ (J - \frac {1} {2} J_{cc})q^2
\end{array} 
\right\}
\end{equation}
All four contributions are functions of the internuclear separation $R$ since all the TB parameters except $\Delta_0$ on the right-hand sides are distance dependent. The covalent and ionic contributions are also explicitly dependent on the charge $q$.

\mbox {} \\ We see that the non-orthogonality of the atomic orbitals enters only the repulsive overlap and attractive covalent-bond contributions explicitly.  The former represents the well-known quantum mechanical overlap repulsion, whereas $\beta_0$ in the latter is the usual Slater-Koster bond-integral modified by the shift $\frac {1} {2} (\alpha_B + \alpha_A) S$  through Eq. (58).  The origin of both these overlap terms can be understood by considering the homovalent s-valent dimer with $q = 0$. In this case, the non-orthogonal TB secular Eq. (8) has matrix elements

\begin{equation}
H_{AA} = H_{BB} = E^0 + \alpha
\end{equation}
from Eq. (44) and

\begin{equation}
H_{AB} = \beta_{SK} + E^0 S
\end{equation}
from Eq. (52), where we have made the common TB assumption that $V_{AB} = V_A + V_B$ by neglecting the small covalent-bond contribution to the dimer potential.

\mbox {} \\ These matrix elements lead to the eigenvalues

\begin{equation}
E^{\pm} = E^0 + (\alpha \pm \beta_{SK}) / (1 \pm S) \,.
\end{equation}
These can be written in the more familiar form as 

\begin{equation}
E^{\pm} = E^0 + \alpha^* \pm \beta_0 
\end{equation}
by defining an effective crystal field term

\begin{equation}
\alpha^* = \frac {1} {2} (E^+ + E^-) - E^0 = (\alpha - \beta_{SK} S) / (1 - S^2)
\end{equation}
and bond integral

\begin{equation}
\beta^* = \frac {1} {2} (E^+ - E^-) = (\beta_{SK} - \alpha S) / (1 - S^2) \,.
\end{equation}
It follows from Eq. (58) that $\alpha^* = (\alpha - \beta_0 S)$ and $\beta^* = \beta_0$ to first order in $S$. Thus, we see that within the total energy expression, Eq. (83), the shift in the eigenvalues $\alpha^*$  enters the overlap repulsion through $2|\beta_0|S$ in Eq. (85) and the electrostatic contribution through $(\alpha_B + \alpha_A)$ in Eq. (86) (for the 2 valence electrons per dimer). On the other hand, the bond integral $\beta_0$ enters the covalent bond contribution as $2|\beta_0|\Theta$.

\mbox {} \\ The electrostatic contribution, Eq. (86), comprises the Coulomb energy of overlapping frozen neutral atoms (the bracketed term) and the Coulomb self-energy of the neutral covalent bond.  This is complemented by the ionic contribution, Eq. (85), that accounts for the flow of charge $q$ that is initially driven by the atomic energy-level mismatch $\Delta_0$. Whereas this leads to the first-order contribution $- \Delta_0 q$ in Equation (88), the other first-order contributions $- (J_{BB} - J_{AA})q$  and $- (J_{Bc} - J_{Ac})\sqrt {1 - q^2}$  in the band energy, Eq. (79), are cancelled by corresponding terms in the double-counting energy, Eq. (82).  These had arisen in the band contribution from the shift in the average on-site energy in Eq. (54) due to the charge $q$ and bond order $\Theta = \sqrt {1 - q^2}$ correcting the dimer potential beyond that of overlapping neutral frozen atoms.  This important result is an example of the Harris-Foulkes DFT functional that is accurate to first order even though its input potential is that due to overlapping frozen atoms (c.f. section 3.4 of [33]).

\mbox {} \\ For a given internuclear separation the total energy will be stationary for the value of q that satisfies

\begin{equation}
\frac {\partial U (R,q)} {\partial q} = O \,.
\end{equation}
It is trivial to show that this stationary condition for the total energy leads to the same self-consistency equation for $q$ that was obtained by solving the TB secular equation self-consistently, namely Eq. (61).  This confirms that we have indeed solved the AB $s$-valent dimer problem correctly within our assumptions of a minimal basis and neglect of explicit exchange-correlation terms.  Just as this self-consistency equation, Eq. (61), was simplified to Eq. (62) by making well-justified approximations, so traditional TB approximates Eqs. (85) - (88) by neglecting the small contributions arising from the differences in the crystal fields, $(\alpha_A^B - \alpha_A^A)$ and the Coulomb integrals involving the covalent-bond density $\rho_{cov}$ .  That is, the total energy is given by Eq. (83) with the electrostatic, covalent and ionic contributions approximated by

\begin{equation}
U_{es} (R) = 1 / R + \alpha_B + \alpha_A  - J_{AB} \,,
\end{equation}

\begin{equation}
U_{cov} (R,q) = - 2 |\beta_0| \sqrt {1 - q^2} \,,
\end{equation}
and

\begin{equation}
U_{ionic} (R,q) = - \Delta_0 q + J q^2 \,.
\end{equation}
Finally, the force is given by 

\begin{align}
F &= - \frac {d U [R,q (R)]} {dR}  \\ \nonumber
&= - \frac {\partial U} {\partial R} - \frac {\partial U} {\partial q} \frac {d q} {d R} = - \frac {\partial U} {\partial R} \,,
\end{align}
because the second term vanishes by Eq. (95). Thus, 

\begin{equation}
F = 
\left\{ 
\begin{array}{l}
- [U_{over}^{'} (R) + U_{es}^{'} (R)] \\
- 2 |\beta_O^{'} (R) | \sqrt {1 - q^2} - J_{AB}^{'} (R)q^2
\end{array} 
\right\}
\,,
\end{equation}
where $g^{'} = dg /dR$ and we have used the approximate expressions (96) - (98). This is consistent with the Hellmann-Feynman theorem in that the force does not depend on the derivative of the density matrix elements (sections 3.1 and 7.6 of [3]). From Eqs. (76) - (78) this is equivalent to the force being independent of either the first-order change in the charge q or the bond order $\Theta = \sqrt {1 - q^2}$.

\subsection {Free-atom limit}
In the limit as the internuclear separation tends to infinity the total energy in Eq (83) tends to the free-atom limit

\begin{equation}
U (R \rightarrow \infty, q) = (U_A^0 + U_B^0) - \Delta_0 q + \frac {1} {2} (J_{BB} + J_{AA}) q^2 \,.
\end{equation}
This corresponds to the energy of free atoms A and B with net charges $Q_A = - q$ and $Q_B = q$ respectively, where

\begin{equation}
\left. 
\begin{array}{l}
U_A (Q) = U_A^0 + |E_A^0| Q + \frac {1} {2} J_{AA} Q^2 \\
U_B (Q) = U_B^0 + |E_B^0| Q + \frac {1} {2} J_{BB} Q^2
\end{array} 
\right\}
\,.
\end{equation}
This second-order expansion of the free-atom energy in terms of the net charge $Q$ allows the TB linear prefactor $|E_{A (B)}^0|$ and quadratic prefactor $\frac {1} {2} J_{AA (BB)}$ to be identified with the experimental ionization potential (IP) and electron affinity (EA) of the respective atoms.  The latter are defined by 

\begin{equation}
IP = U (Q=1) - U (Q=0) 
\end{equation}
and
\begin{equation}
EA = U (Q=0) - U (Q=-1) \,, 
\end{equation}
where U(Q) is the energy of the charged atom.  It follows from Eqs. (102) - (104) that

\begin{equation}
|E^0| = \frac {1} {2} (IP + EA) = \chi^0  
\end{equation}
and

\begin{equation}
\frac {1} {2} J^0 = \frac {1} {2} (IP - EA) = \eta  
\end{equation}
with $J^0 = J_{AA (BB)}$. $\chi^0 $ is the Mulliken {\it electronegativity} which he defined as the average of the ionization potential and electron affinity of the free atoms (page 107 [43]). $\eta$ is called the {\it chemical hardness} of the free atom and is defined by the quadratic prefactor in the free-atom energy expansion, Eq. (102) (page 107 [4]).  Thus, within this simple s valent TB model the magnitude of the energy level of the neutral free atom plays the role of electronegativity, whereas one-half of the on-site Coulomb integral (or Hubbard U) plays the role of hardness. 

\mbox {} \\ The change in total energy with the flow of charge can be written from Eq. (101) in the free-atom limit as

\begin{equation}
\Delta U (R \rightarrow \infty, q) = - (\Delta \chi^0) q + 2 \eta q^2   
\end{equation}
where $\Delta \chi^0 = \chi^0_B - \chi^0_A$ and $\bar{\eta} = \frac {1} {2} (\eta_B + \eta_A)$ .  The values of $\chi^0$ and $\eta$ for H and Li are given in table 1.
\\
\\

\begin{table}[htpb]
\begin{center}
\begin{tabular}{|c|c|c|c|c|} \hline
& \bf IP (eV) & \bf EA (eV) & $\chi^0 = |E^0| = \frac {1} {2} (IP + EA)$ & $\eta = \frac {1} {2} J^0 = \frac {1} {2} (IP - EA)$ \\ \hline
\bf H & 13.6 & 0.8 & 7.2 & 6.4 \\ \hline
\bf Li & 5.4 & 0.6 & 3.0 & 2.4 \\ \hline

\end{tabular}
\end{center}
\caption{\label{tab:1}  Experimental values of ionization potential, electron affinity, and corresponding values of electronegativity (magnitude of free-atom energy level) and chemical hardness (one-half on-site Coulomb integral) for hydrogen and lithium.}
\end{table}

\mbox {} \\ Fig. 5 shows the resultant energy curve $\Delta U (R \rightarrow \infty, q)$ for LiH with $\Delta \chi^0 = 4.2$ eV and $\bar{\eta} = 4.4$ eV at infinite internuclear separation. We see that charge is driven from the Li atom to the H atom by the attractive linear term proportional to their electronegativity difference $\Delta \chi^0$ , but that this is countered by the repulsive quadratic term proportional to their average chemical hardness $\bar{\eta}$.  Provided $\Delta \chi^0 / 4 \bar{\eta} \leq 1$ the total energy, Eq. (107) is stationary for 

\begin{equation}
q_{\infty} = \frac {\Delta \chi^0} {4 \bar{\eta}} = \frac {\Delta_0} {J_{AA} + J_{BB}} \,,  
\end{equation}
when the binding energy with respect to the neutral free atoms takes the Pauling form for the heat of formation [5], namely

\begin{equation}
\Delta H = - \frac {1} {8 \bar{\eta}} (\Delta \chi^0)^2 \,.  
\end{equation}

\mbox {} \\ These last two expressions give $q_{\infty} = 0.24$ and $\Delta H = - 0.50$ eV for the case of LiH in the free-atom limit.
In reality when the Li and H atoms are separated to infinity, the isolated atoms cannot contain fractional charges since there will be no hopping or tunnelling of the electrons between the sites.  The only physical configurations would be $Li^0 H^0, Li^+ H^-$ and $Li^- H^+$.  But from Fig. 5 we see that the configuration $Li^+ H^-$ costs nearly 5eV in energy compared to the charge neutral state, and $Li^- H^+$ would still be a further 8.4eV higher in energy.  Thus, energetically nature favours the separation of the LiH dimer into neutral atoms.  The prediction of fractional charges in the free-atom limit is a well-recognized problem with TB and DFT because they have not imposed any constraint on integer numbers of electrons.  We should, therefore, be aware that these one-electron-type methods can lead to an over-estimation of charge flow at large internuclear separations (Eq. (46) of [4]).

\begin{figure} 
\center
\includegraphics[width=.5\textwidth]{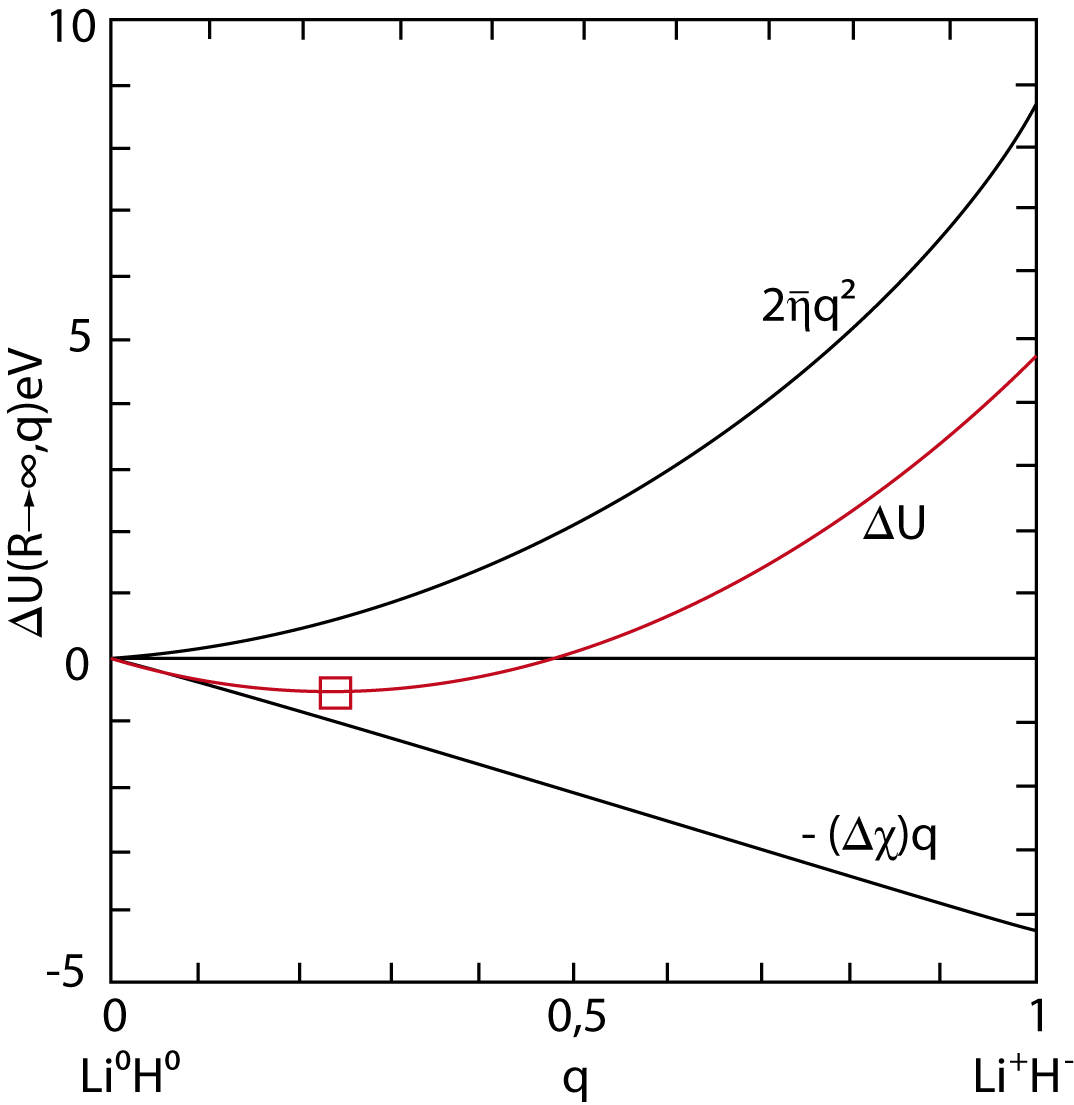}
\caption{Changes in total energy of LiH in limit of infinite internuclear separation as function of net charge q on Li atom. $\Delta \chi =|E_{Li}^0 - E_H^0|$ is electronegativity difference and $\bar{\eta} = \frac {1} {4}(J_{LiLi} + J_{HH})$ is average chemical hardness. Open square marks position where energy is stationary at $q_{\infty}=0.24$.} \label{fig:LiH} 
\end{figure}

\subsection {Binding energy}
The {\it binding} energy of the AB dimer is obtained by subtracting off the energy of the {\it neutral} free atoms A and B, Eq. (84), from the total energy, Eq. (83).  That is, 

\begin{equation}
U_{be} (R,q) = U_{rep} (R) + U_{cov} (R,q) + U_{ionic} (R,q) \,,  
\end{equation}
where the repulsive contribution is given by

\begin{equation}
U_{rep} (R) = U_{over} (R) + U_{es} (R) \,.  
\end{equation}
We will make the customary TB approximation as in Eqs. (96) - (98) to write

\begin{equation}
U_{cov} (R,q) = - 2|\beta_0 (R)|\sqrt {1 - q^2} \,,  
\end{equation}
and

\begin{equation}
U_{ionic} (R,q) = - \Delta_0 q + J (R) q^2 \,.   
\end{equation}
As we have seen in the previous section this binding-energy expression tends to the wrong limit, Eq. (108), as the atoms are pulled infinitely apart. However, this need not be of a major concern to us as TB and DFT are known to give an accurate representation of the energetics of most materials in the vicinity of their equilibrium separations or atomic volumes, where the effective one-electron-type description works well.

\mbox {} \\ Let us consider first the homovalent hydrogen molecule since the ionic contribution vanishes as $\Delta_0 = 0$. The upper panel of Fig. 6 shows the bonding and anti-bonding energy levels that are predicted by non-spin polarized DFT.  This neglects the occurrence of local magnetic moments that are stable for internuclear separations greater than about $2R_0$, where $R_0$ = 1.4 au is the equilibrium distance (see, for example, Fig. 3.6 of [1]). 

\begin{figure} 
\center
\includegraphics[width=.7\textwidth]{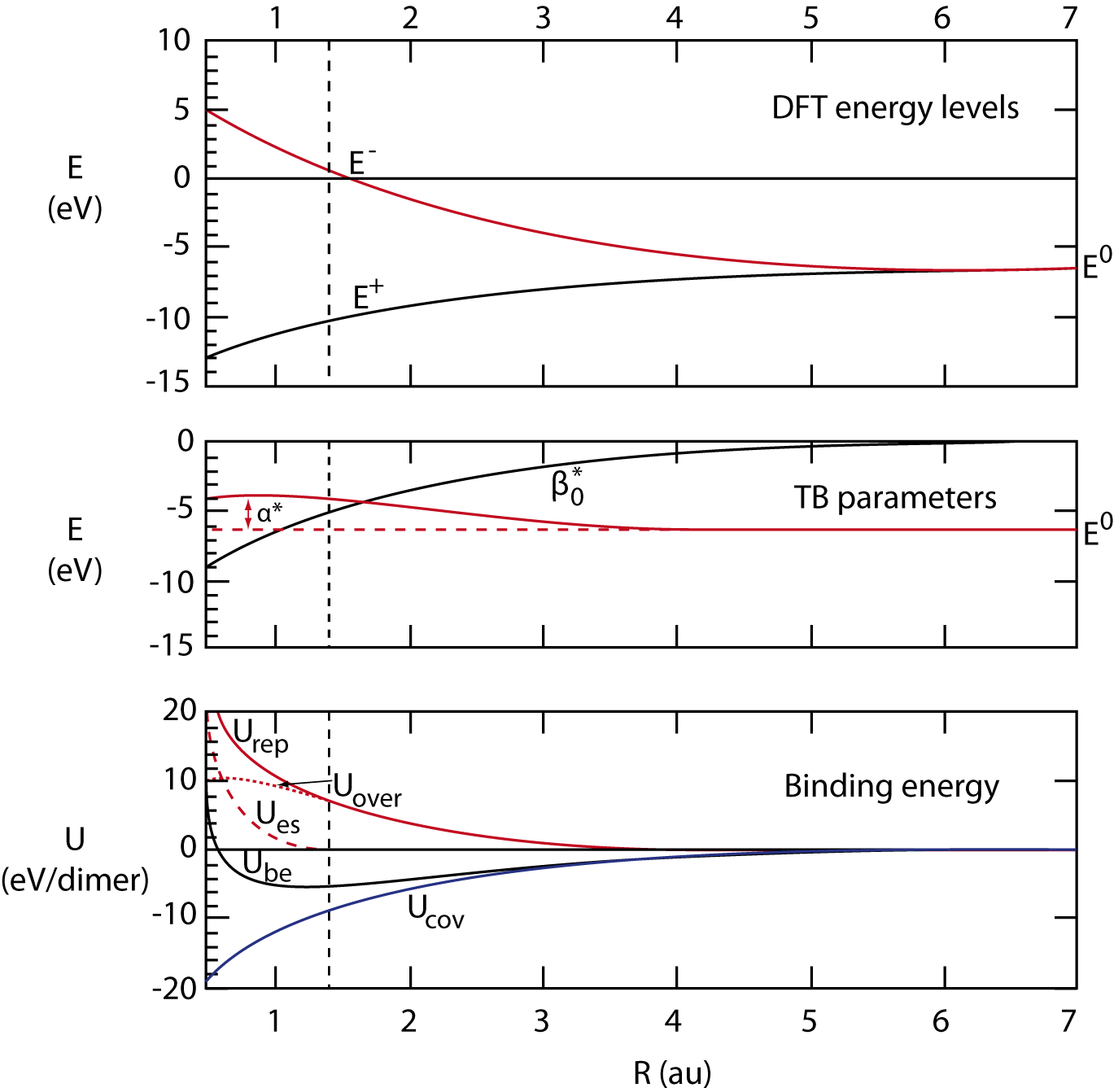}
\caption{TB fit to DFT energy levels and binding energy curve as function of internuclear separation $R$ for $H_2$ with equilibrium distance marked by dashed vertical line [6].  Upper panel: DFT bonding $(E^+)$ and anti-bonding $(E^-)$ energy levels.  Middle panel: TB parameters $E^0, \alpha^*$ and $\beta_0$ from fitting DFT data in upper panel.  Lower panel: DFT binding energy (solid black curve), covalent-bond energy $- 2 |\beta_0|$ (solid blue curve), repulsive energy (solid red curve) with its overlap (dotted red curve) and electrostatic (dashed red curve) components.} \label{fig:TB_to_DFT} 
\end{figure} 

\mbox {} \\ These two DFT eigenvalues can be used to fit the three TB parameters $E^0, \alpha^* (R)$ and $\beta_0 (R)$ because both $\alpha^* (R)$  and $\beta_0 (R)$  tend to zero at large internuclear separations, as shown in the middle panel of Fig. 6.  We see that the repulsive overlap contribution $|\beta_0|S$  in Eq. (93) dominates over the attractive crystal field term $\alpha$ in Eq. (92) as $\alpha^* > 0$ over the range of distances plotted.  However, we observe that for $R \leq R_0 \alpha^*$ starts to saturate as the third order attractive contribution $\alpha^* S^2$ in Eq. (93) becomes important for $S \geq \frac {1} {2}$.  At the equilibrium seperation $\alpha^* = 2.0$eV and $\beta_0 = 5.8$eV.

\mbox {} \\ Having used the DFT energy levels to fit the TB parameters $E^0, \alpha^*$ and $\beta_0$, we can now use the DFT binding energy curve in the lower panel of Fig. 6 to extract the repulsive energy, from which we may deduce the corresponding overlap and electrostatic contributions. The repulsive energy in Eq. (111) can be expressed as

\begin{equation}
U_{rep} (R) = U_{be} (R) - U_{cov} (R) \,,   
\end{equation}
where $U_{cov} (R) = 2 |\beta_0 (R)|$ from Eq. (112) as the hydrogen molecule has a fully saturated bond with a bond order of unity.  Subtracting the resultant covalent-bond contribution in Fig. 6 from the binding energy we recover the repulsive curve that is plotted.

\mbox {} \\ This repulsive contribution may be separated into its overlap and electrostatic constituents as follows.  Firstly, the overlap term can be deduced by assuming that the overlap integral is approximately proportional to the bond integral, so that the overlap repulsion varies as

\begin{equation}
U_{over} (R) = A[\beta_0 (R)]^{\lambda} \,,   
\end{equation}
The dotted curve in Fig. 6 demonstrates an extremely good fit to the DFT-derived repulsion for $R \geq R_0$  with A = 0.185 and $\lambda$ = 1.85 (giving the overlap energy in Eq. (115) in units of eV/dimer provided $\beta_0$  is in eV).  This value of 1.85 is remarkably close to the original Wolfsberg-Helmholz approximation of $\lambda$ =2 [1].
Secondly, the electrostatic contribution can be obtained by subtracting the overlap repulsion from the DFT-deduced repulsive curve in Fig. 6.  The resultant dashed curve kicks-in for $R \leq R_0$ and has a shape that agrees with that obtained analytically by evaluating the electrostatic interaction between neutral atoms comprising 1s atomic charges resulting from variational hydrogonic wave functions $(\zeta^3)^{\frac {1}{2}}$ exp $(- \zeta r)$  (c.f. Eq. (32) and Fig. 6 of [7]). Traditional TB methods write the repulsive energy as a sum over pair potentials, namely

\begin{equation}
U_{rep} (R) = \frac {1}{2} \sum_{i,j} \Phi_{rep} (R) \,.   
\end{equation}
That is, for the dimer $U_{rep} (R) = \Phi_{rep} (R)$ This pairwise assumption is justified to lowest order in bulk materials: the electrostatic interaction is pairwise within the usual TB approximation that the bulk potential is the sum over atomic potentials and the overlap contribution, resulting from the non-orthogonality shifts in the on-site energies, is also pairwise as $\sum_{i,j} H_{ij} S_{ji}$ to first order in $S$.

\mbox {} \\ We see, therefore, that for {\it homovalent} dimers the binding energy can be decomposed in terms of the physically-based contributions of overlap repulsion, atom-atom electrostatic interaction, and covalent bond energy.  The firm physical basis of this TB model is confirmed by the fact that the bond integrals extracted from the DFT eigenvalues leads to a covalent-bond contribution to the energy that reproduces the DFT binding energy curves almost exactly for $R > 2R_0$, when at smaller distances the repulsive overlap energy becomes noticeable.  Further, the electrostatic atom-atom interaction is found to provide the rapid increase in repulsion at short distances.  This soundness of the TB model underpins its application to many material systems where it is assumed that the two-centre bond integrals $\beta_0 (R)$ and the pairwise repulsion $\Phi_{rep} (R)$ are transferable from one crystal structure or atomic configuration to another.

\mbox {} \\ Finally, we consider the {\it heterovalent} dimer LiH  where the ionic contribution also enters the TB expression for the binding energy, Eq. (110).  We have seen from Eq. (62) that the self-consistent charge q is a function of the normalized bond integral $\tilde {\beta}_0 = \beta_0 / 2 J$ and atomic energy-level mismatch $\tilde {\Delta}_0 = \Delta_0 / 2 J$ , where $J = \frac {1}{2} (J_{BB} + J_{AA}) - J_{AB} (R)$. From table 1 the values of $\Delta_0 = E^0_B - E^0_A  = 4.2$eV and $\frac {1}{2} (J_{HH} + J_{LiLi}) = 8.8$eV.  The value of the intersite Coulomb integral $J_{AB} (R)$ can be found from the interpolation formula [8]

\begin{equation}
J_{AB} (R) = \frac {1}{[R^3 + (1 / \sqrt {J_{AA} J_{BB}})^3]^{\frac {1}{3}}} \,   
\end{equation}
using atomic units with $R$ in au and the Coulomb integrals in Hartrees.  For large distances when the A and B atoms do not overlap this falls off Coulombically as $1/R$. However, for short distances it saturates as the atomic orbitals overlap.  For R = 0 it takes the value $\sqrt {J_{AA} J_{BB}}$ which is exact for the homovalent case when the A and B atoms are chemically identical.  From table 1 for LiH $\sqrt {J_{AA} J_{BB}} = 7.8$eV$ = 0.29$ Hartree.
Thus, 

\begin{equation}
J_{LiH} (R) = \frac {27.2}{[R^3 + 41.9]^{\frac {1}{3}}} \, \text{eV} \,.   
\end{equation}
This is plotted in Fig. 7 where we see that the saturation from the unscreened Coulomb inverse power law becomes very important for $R \leq 5$ au.

\begin{figure} 
\center
\includegraphics[width=.5\textwidth]{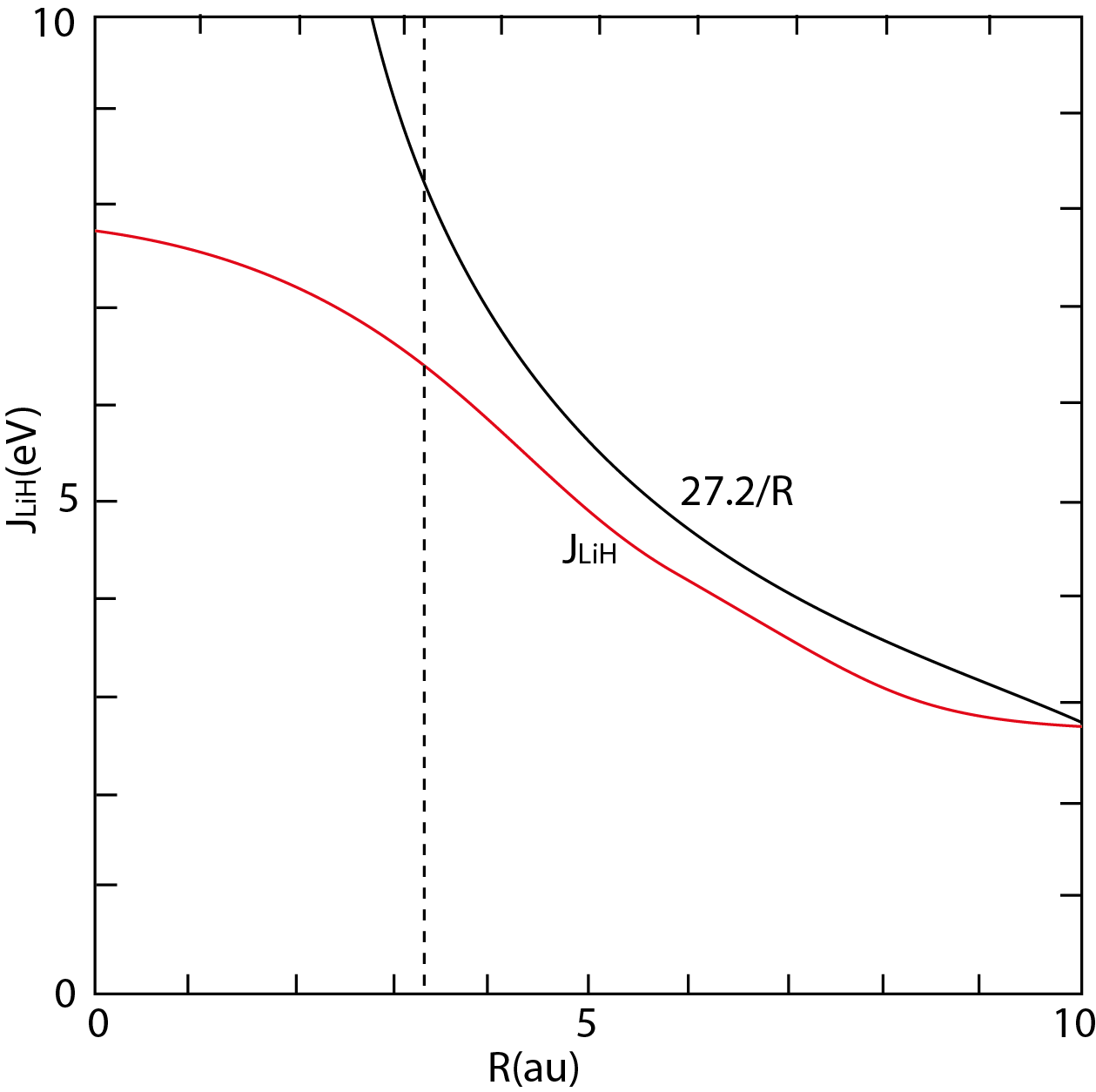}
\caption{Comparison of LiH interatomic Coulomb integral $J_{LiH}(R)$ with unscreened Coulomb inverse power law for a wide range of internuclear seperations R.  Vertical dashed line gives equilibrium internuclear separation.} \label{fig:JLiH} 
\end{figure}

\mbox {} \\ Table 2 gives the DFT values of the equilibrium internuclear separation $R_0$ and bond integral $\beta_0$ for H$_2$ and Li$_2$. The equilibrium separation for LiH was taken from Zen's law as the arithmetic mean of the equilibrium separations for Li$_2$ and H$_2$.  The equilibrium bond integral for LiH, on the other hand, was assumed to be the geometric mean of the equilibrium bond integrals for Li$_2$ and H$_2$ (as for the LiH on-site Coulomb integral in Eq. (117)). We see from Fig. 7 that the equilibrium value of J$_{LiH}$ is then 6.4eV.  Hence, J in Eq. (62) takes the value (8.8 - 6.4) = 2.4eV, so that the presence of the attractive nearest neighbour Coulomb interaction dramatically reduces the effective on-site Coulomb integral, thereby enhancing the propensity for charge flow.  Thus, LiH at equilibrium takes the co-ordinate point $(2 \tilde {\beta}_0, \tilde {\Delta}_0) $ = (1.20, 0.88) on the charge contour map in Fig. 4.  This corresponds to a value of $q = 0.4$, so that, as expected, the ionic and covalent bonds in LiH play an almost equal role.

\begin{table}[htpb]
\begin{center}
\begin{tabular}{|c|c|c|c|} \hline
& \bf H$_2$ & \bf Li$_2$ & \bf LiH \\ \hline
$R_0 $(au) & 1.4 & 5.1 & 3.3 \\ \hline
$\beta_0 (R_0) $(eV) & 5.8 & 1.5 & 2.9 \\ \hline
\end{tabular}
\end{center}
\caption{\label{tab:2}  Equilibrium bond lengths and bond integrals for H$_2$, Li$_2$ and LiH.}
\end{table}

\section {Conclusions}

These notes have shown how it is possible to derive rigorously a TB expression for the binding energy of an AB $s$-valent dimer by starting from the effective one-electron-type equations of DFT.  The TB binding energy has four well-defined contributions that are physically and chemically motivated, namely the overlap repulsion, the atom-atom electrostatic interaction, and the attractive covalent and ionic bond energies respectively.  This TB binding energy in its turn can then be used to justify well-founded ionic or covalent interatomic potentials.  Hopefully these notes will help you the reader to hone your own physical intuition about the nature of chemical bonding in materials, and hence to choose your own path through the myriad of different TB approaches out there in the literature. \\[0,5cm]

\section {Acknowledgments}
I would like to thank Christa Hermichen for producing such excellent digital figures.

\mbox {} \\ {\bf References (plus references therein)}

\begin{enumerate}
\item	D.G. Pettifor, Bonding and Structure of Molecules and Solids, OUP (1995).

\item	J.K. Burdett, Chemical Bonding in Solids, OUP (1995).

\item	M.W. Finnis, Interatomic Forces in Condensed Matter, OUP (2003).

\item	S.W. Rick and S.J. Stuart, Potentials and Algorithms for Incorporating Polarizability in Computer Simulations, Rev. Comp. Chem. 18 (2002) 89.

\item	L. Pauling, The Nature of the Chemical Bond, Cornell University, Ithaca (1960).

\item	R. Drautz and D.G. Pettifor, unpublished (2004).

\item	A.J. Skinner and D.G. Pettifor, Transferability and the Pair Potential within the Tight-Binding Bond Model: an Analytic Study for Hydrogen, J. Phys: Condens. Matter 3 (1991) 2029.

\item	S.L. Njo, J. Fan and B. van de Graaf, Extending and Simplifying the Electronegativity Equalization Method, J. Mol. Catalysis A: Chemical 134 (1998) 79.
\end{enumerate}

\end{document}